\documentclass[a4paper, reprint, twoside]{revtex4-1}

\usepackage[english]{babel}
\usepackage[utf8]{inputenc}
\usepackage[labelfont=up]{subcaption}
\usepackage[x11names]{xcolor}
\usepackage{float}
\usepackage{mathtools}
\usepackage[top=1.in, bottom=1.in, left=0.5in, right=0.5in]{geometry}
\usepackage{amsthm, amssymb}
\usepackage{graphicx}
\usepackage{lipsum}
\usepackage{bm}
\usepackage{wasysym}
\usepackage{enumitem}

\let\vec\bm

\DeclarePairedDelimiter\ton{(}{)}
\DeclarePairedDelimiter\qua{[}{]}

\begin{document}

\title{Tolerance versus synaptic noise in dense associative memories}
\author{Elena Agliari$^1$}
\email[Correspondence email address: ]{agliari@mat.uniroma1.it}
\author{Giordano De Marzo$^2$}
\affiliation{$^1$Dipartimento di Matematica “Guido Castelnuovo”, Sapienza Università di Roma, Roma, Italy.\\ 
$^2$Dipartimento di Fisica, Sapienza Università di Roma, Roma, Italy.}

\date{\today} 
\begin{abstract}
The retrieval capabilities of associative neural networks can be impaired by different kinds of noise: the fast noise (which makes neurons more prone to failure), the slow noise (stemming from interference among stored memories), and synaptic noise (due to possible flaws during the learning or the storing stage). In this work we consider dense associative neural networks, where neurons can interact in $p$-plets, in the absence of fast noise, and we investigate the interplay of slow and synaptic noise. In particular, leveraging on the duality between associative neural networks and restricted Boltzmann machines, we analyze the effect of corrupted information, imperfect learning and storing errors. For $p=2$ (corresponding to the Hopfield model) any source of synaptic noise breaks-down retrieval if the number of memories $K$ scales as the network size. For $p>2$, in the relatively low-load regime $K \sim N$, synaptic noise is tolerated up to a certain bound, depending on the density of the structure.
\end{abstract}

\maketitle

\section{Introduction} \label{sec:intro}

Associative memories (AM) are devices able to \emph{store} and then \emph{retrieve} a set of information (see e.g., \cite{intro}). 
Since the 70's, several models of AM have been introduced, among which the Hopfield neural network (HNN) probably constitutes the best known example \cite{Amit, Coolen}. In this model one has $N$ units, meant as stylized (on/off) neurons, able to process information through pairwise interactions. The performance of an AM is usually measured as the ratio $\alpha$ between the largest extent of information safely retrievable and the amount of neurons employed for this task; in the HNN this ratio is order of $1$.
In the last decades many efforts have been spent trying to raise this ratio (see e.g., \cite{FAB-2019,AABF-2019} and references therein). For instance, in the so-called dense associative memories (DAMs)  neurons are embedded on hypergraphs in such a way that they are allowed to
interact in $p$-plets and $\alpha \sim \mathcal O(N^{p-1})$. However, this model also requires more resources as the number of connections encoding the learned information scales as $N^{p}$ instead of $N^{2}$ as in the standard pairwise model \cite{Krotov,Baldi}. 

Clearly, whatever the AM model considered, limitations on $\alpha$ are intrinsic given that the amount of resources (in terms of number of neurons and number of connections) available necessarily yields to bounds in the extent of information storable. 
In particular, by increasing the pieces of information to be stored, the \emph{interference} among them generates a so-called \emph{slow noise} which requires a relatively large number of neurons or of connections to be resolved. 
Beyond this, one has also to face another kind of noise, which has been less investigated in the last years and which is the focus of the current work. 

In fact, classical AM models assume that communication among neurons is perfect and that learning and storing stages can rely on exact knowledge of information whereas, in general, communication can be disturbed and the information provided may be affected
by some source of noise (see e.g., \cite{AABCF-PRL2020,BM-PRL2020}). We refer to the noise stemming from this kind of flaws as \emph{synaptic noise} and we envisage different ways to model it, mimicking different physical situations. In each case we investigate the effects of such a noise on the retrieval capabilities of the system and on the existence of bounds on the amount of noise above which the network can not work as an AM any longer.
More precisely, our analysis is led on hyper-graphs with $p \geq 2$ and we highlight an interplay between slow noise, synaptic noise and network density: by increasing $p$ one can exploit some of the additional resources to soften the effect of slow noise and make higher load affordable, and some to soften the effect of synaptic noise and make the system more robust.   
On the other hand, here, possible effects due to fast noise (also referred to as temperature) are discarded and, since it typically reduces tolerance, our results provide an upper bound for the system tolerance. Also, this particular setting allows addressing the problem analytically via a signal-to-noise approach \cite{Amit}.

In the following Sec.~\ref{sec:defs}, we will frame the problem more quantitatively exploiting, as a reference model, the HNN: we will review the signal-to-noise approach and introduce the necessary definitions. 
Next, in Sec.~\ref{sec:DAM}, we will consider the $p$-neuron Hopfield model and we will find that $i.$ when the information to be stored is provided with some mistakes, then the machine will store the defective pieces of information and retrieving the correct ones is possible as long as mistakes are ``small''; $ii.$ when the information is provided exactly but the learning process is imperfect, then retrieval is possible but the capacity $\alpha$ turns out to be downsized; $iii.$ when information is provided exactly and it is correctly learned, but communication among neurons during retrieval is impaired, then retrieval is still possible but $\alpha$ is ``moderately'' reduced. These results are also successfully checked versus numerical simulations.
Finally, Sec.~\ref{sec:conclusions} is left for our conclusive remarks.
Since calculations for the $p$-neuron Hopfield model are pretty lengthy, they are not shown in details for arbitrary $p$, instead, we report explicit calculations for the case $p=4$ in the Appendix.


\section{Noise tolerance} \label{sec:defs}
In this section we introduce the main players of our investigations taking advantage of the HNN as a reference framework. \\
The HNN is made of $N$ neurons, each associated to a variable $\sigma_i \in \{-1, +1 \}$, with $i=1, ..., N$ representing the related status (either active or inactive),  embedded in a complete graph with weighted connections. 
An HNN with $N$ neurons is able to learn pieces of information which can be encoded in binary vectors of length $N$, also called patterns. After the learning of $K$ such vectors $\{ \boldsymbol \xi^1, ..., \boldsymbol \xi^K \}$, with $\boldsymbol \xi^{\mu} \in \{-1, +1\}^K$ for $\mu=1,...,K$, the weight for the coupling between neuron $i$ and $j$ is given by the so-called Hebbian rule $J^{Hebb}_{ij}= \frac{1}{N} \sum_{\mu=1}^K \xi_i^{\mu} \xi_j^{\mu}$ for any $i \neq j$, while self-interactions are not allowed, i.e., $J_{ii}=0$, for any $i$.
\newline
In the absence of external noise and external fields, the neuronal state evolves according to the dynamic
\begin{equation} \label{eq:dynamic}
\sigma_i(t+1) = \textrm{sign}[h_i(\boldsymbol \sigma(t))],
\end{equation}
where
\begin{equation} \label{eq:field}
h_i(\boldsymbol \sigma(t)) = \sum_{j=1}^N J_{ij} \sigma_j(t)
\end{equation}
is the internal field acting on the $i$-th neuron.
This dynamical system corresponds to a steepest descent algorithm where
\begin{equation} \label{eq:hopfield}
H(\boldsymbol \sigma, \boldsymbol \xi) = \sum_{i>j}^N h_i(\boldsymbol \sigma) \sigma_ i =  \frac{1}{2N} \sum_{\substack{i,j \\ i \neq j}}^{N,N} \sum_{\mu=1}^K \xi_i^{\mu} \sigma_i \sigma_j \xi_j^{\mu}
\end{equation}
plays as a Lyapunov function or, in a statistical-mechanics setting, as the Hamiltonian of the model (see e.g., \cite{Amit, Coolen}).\\ 
The retrieval of a learned pattern $\boldsymbol \xi^{\mu}$, starting from a certain input state $\boldsymbol \sigma (t=0)$, is therefore assured as long as this initial state belongs to the attraction basin of  $\boldsymbol \xi^{\mu}$, according to the dynamic (\ref{eq:dynamic}), in such a way that, eventually, the neuronal configuration will reach the stable state $\boldsymbol \sigma = \boldsymbol \xi^{\mu}$.
With these premisis, the signal-to-noise analysis ascertains the stability of the arbitrary pattern $\boldsymbol \xi^{\mu}$ by checking whether the inequality
\begin{equation} \label{eq:s2n}
h_i(\boldsymbol \xi^{\mu}) \xi_i^{\mu} >0
\end{equation}
is verified for any neuron $i=1,...,N$.
Of course, this kind of analysis can be applied to an arbitray AM model, by suitably defining the internal field in the condition (\ref{eq:s2n}), as $h_i$ issues from the architecture characterizing the considered model.\\
Before proceeding, a few remarks are in order. \\
The expression ``signal-to-noise'' refers to the fact that, as we will see, the l.h.s. in (\ref{eq:s2n}) can be split into a ``signal'' term $S$ and a ``noise'' term $R$, the latter typically stemming from interference among patterns and growing with $K$. Thus, the largest amount of patterns that the system can store and retrieve corresponds to the largest value of $K$ which still ensures $S/R \gtrsim 1$. Further, since we are interested in storing the largest amount of information, rather than the largest amount of patterns, recalling the Shannon-Fano coding, the pattern entries shall be drawn according to 
\begin{equation}
P(\xi_i^{\mu}) = \frac{1}{2}[\delta(\xi_i^{\mu} + 1) + \delta(\xi_i^{\mu} -1)],
\end{equation}
for any $i, \mu$, that is, entries are taken as i.i.d. Rademacher random variables.

Remarkably, the above mentioned Hebbian rule accounts for
perfect $i.$ dataset, $ii.$ learning and $iii.$ storage of information whereas, in general, some source of noise may take place and, according to the stage where it occurs, we revise $J^{Hebb}$ as explained hereafter.
\begin{enumerate}[label=(\roman*)]
\item
\begin{figure}[tb]
			\includegraphics[width=0.33\textwidth]{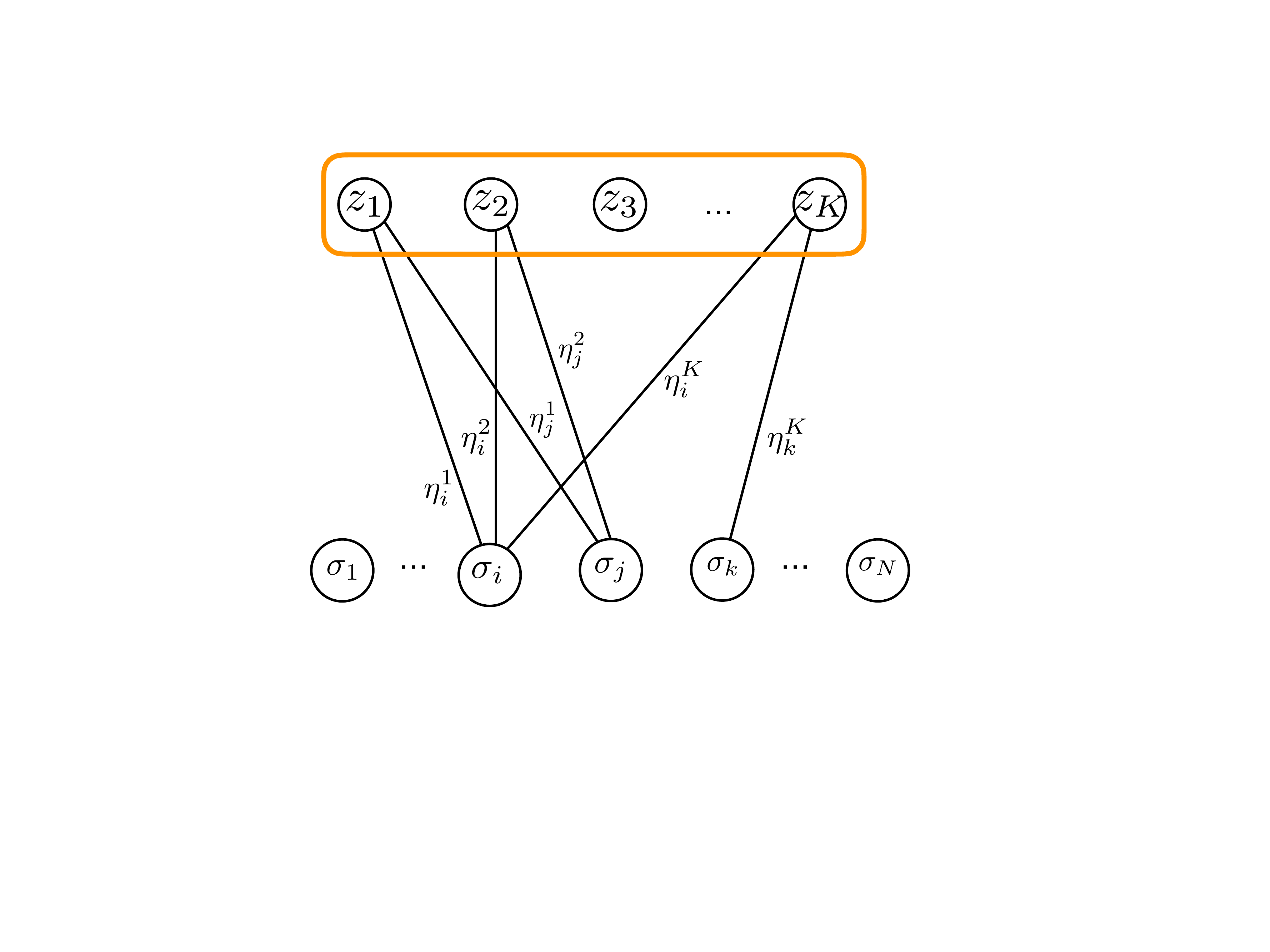}
			\caption{\textbf{RBM corresponding to faulty patterns.} The machine is built over a hidden layer made of Gaussian neurons $\{z_{\mu}\}_{\mu=1,...,K}$ and a visible layer made of binary neurons $\{ \sigma_i \}_{i=1,...,N}$; in this case a neuron $z_{\mu}$ belonging to the hidden layer can interact with one neuron $\sigma_i$ belonging to the visible layer and the coupling is $\eta_i^{\mu} = \xi_i^{\mu} + \omega \tilde{\xi}_i^{\mu}$, as described by Eq.~\ref{eq:noisy_patterns}. Since the machine is restricted, intra-layer interactions are not allowed. In the dual associative network the neurons interact pairwise ($p=2$) and the synaptic weight for the couple ($\sigma_i, \sigma_j$) is $J_{ij} = \sum_{\mu} ( \xi_i^{\mu} + \omega \tilde{\xi}_i^{\mu})( \xi_i^{\mu} + \omega \tilde{\xi}_i^{\mu})$, as reported also in Eq.~\ref{eq:noisy_couplings_a}. This structure can be straightforwardly generalized for $p>2$. In this figure, seeking for clarity, only a few connections are drawn for illustrative purposes.}
\label{fig:map_a}
		\end{figure}
The first kind of noise we look at allows for corrupted patterns, referred to as $\{\boldsymbol \eta^{\mu}\}_{\mu=1,...,K}$, and defined as
\begin{equation} \label{eq:noisy_patterns}
\eta_i^{\mu}=\xi_i^{\mu}+\omega ~ \tilde{\xi}_i^{\mu},
\end{equation}
where  $\tilde{\xi}_i^{\mu}$ is a standard Gaussian random variable and $\omega$ is a real parameter that tunes the noise level.
The Hebbian rule, in the case $p=2$, is therefore revised as 
\begin{equation} \label{eq:noisy_couplings_a}
J_{ij} = \frac{1}{N}\sum_{\mu=1}^K \eta_i^{\mu} \eta_j^{\mu}.
\end{equation}
This is an inner and rather strong kind of noise, in fact, as we will show, even in a low-load regime (i.e., $K/N^{p-1} \to 0$) and for relatively small values of $\omega$, it implies the breakdown of pattern recognition capability. It is intuitive to see that this kind of noise leads to such a dramatic effect if one looks at the dual representation of the associative neural network in terms of a restricted Boltzmann machine (RBM) \cite{AABCF-PRL2020, Bernacchia, ABGGM-PRL2020,ABDG-NN2013,AGST-PRE2017,AGST-PRE2018}, see Fig.~\ref{fig:map_a}. In fact, the coupling (\ref{eq:noisy_couplings_a}) is reminiscent of the fact that, during the learning stage, the system is fed by noisy patterns and therefore it learns patterns along with their noise. Notice that, when neurons interact $p$-wisely, the coupling $J_{ij}$ turns out to be a polynomial order $p$ in $\omega$.
\item
The second kind of noise we look at can be thought of as due to flaws during the learning stage. Still looking at the RBM representation, in this case the couplings between visible and hidden units are noisy and, again, we quantify this noise by $\omega$ times a standard Gaussian variable, see Fig.~\ref{fig:map_b}. Notice that, when $p=2$ (as for the classical HNN), this kind of noise coincides with the previous one and, in general, it yields to a revision in the coupling $J^{Hebb}_{ij}$ given by additional terms up to second order in $\omega$. This suggests that, in this case, effects are milder with respect to the previous one. In fact, as we will see, in a low-load regime, the degree of noise $\omega$ can grow algebraically with the system size, without breaking retrieval capabilities. 
		\begin{figure}
			\includegraphics[width=0.33\textwidth]{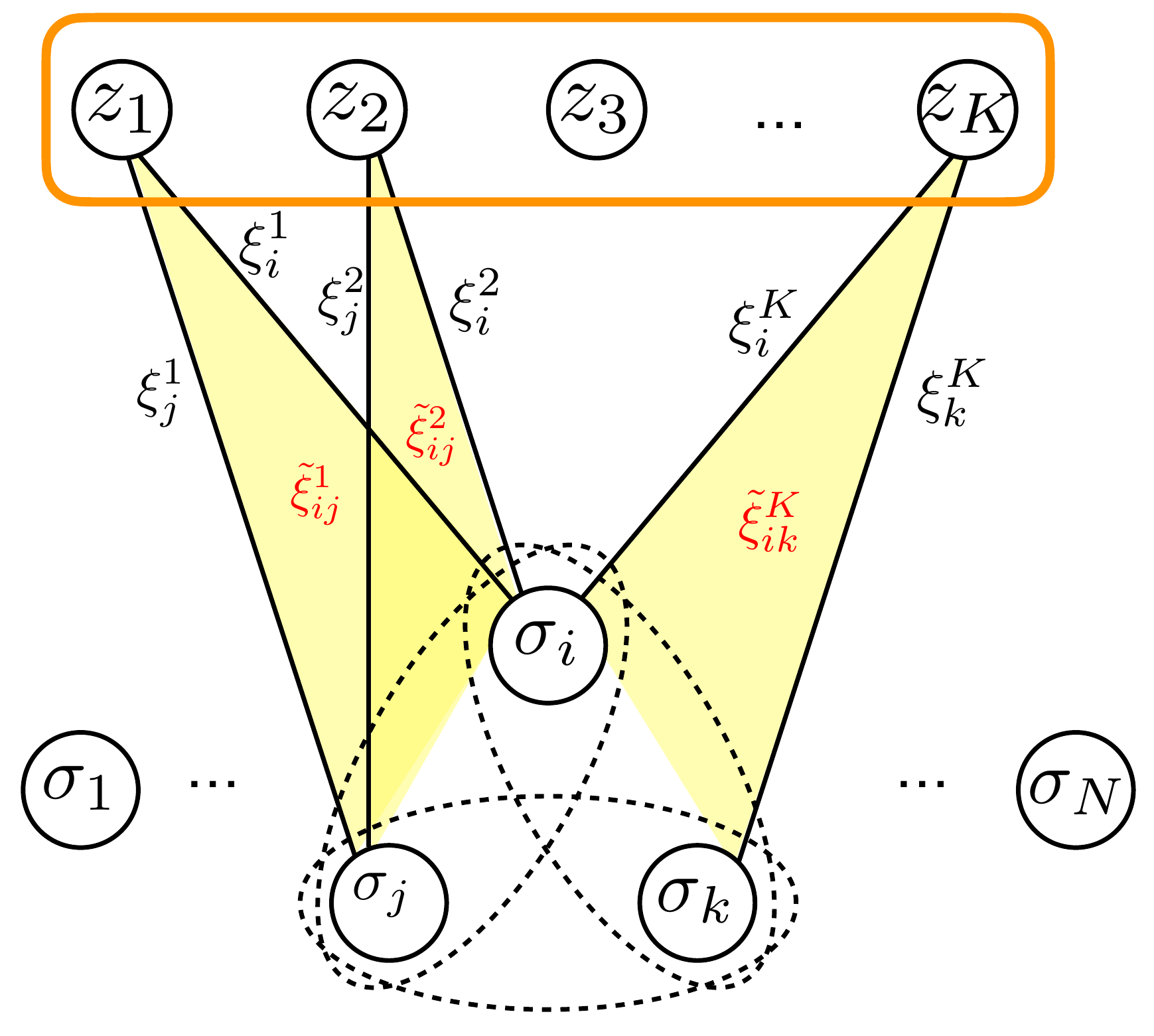}
			\caption{\textbf{RBM corresponding to shortcomings in the learning stage.}  The machine is built over a hidden layer made of Gaussian neurons $\{z_{\mu}\}_{\mu=1,...,K}$ and a visible layer made of binary neurons $\{ \sigma_i \}_{i=1,...,N}$; in this case a neuron $z_{\mu}$ belonging to the hidden layer can interact simultaneously with two neurons $(\sigma_i, \sigma_j)$ belonging to the visible layer and the coupling is $\xi_i^{\mu}\xi_j^{\mu} + \omega \tilde{\xi}_{ij}^{\mu}$, mimicking a situation where the correct patterns are learnt, yet interaction among the two layers is disturbed. Since the machine is restricted, intra-layer interactions are not allowed. In the dual associative network the neurons interact $4$-wise ($p=4$) and the synaptic weight for the $4$-plet ($\sigma_i, \sigma_j, \sigma_k, \sigma_l$) is $J_{ijkl} = \sum_{\mu}  ( \xi_i^{\mu} \xi_j^{\mu} + \omega \tilde{\xi}_{ij}^{\mu})( \xi_k^{\mu} \xi_l^{\mu} + \omega \tilde{\xi}_{kl}^{\mu})$, as reported also in Eq.~\ref{eq:coupling_b}. Notice that this kind of noise is intrinsically defined only for associative networks where $p$ is even and that when $p=2$ we recover case depicted in Fig.~\ref{fig:map_a}. Also in this figure, seeking for clarity, only a few connections are drawn for illustrative purposes.}
\label{fig:map_b}
		\end{figure}
\item
The third kind of noise we look at can be thought of as due to effective shortcomings in storage as it directly affects the coupling among neurons in the AM system as
\begin{equation} \label{eq:noisy_couplings_c}
J_{ij} = \frac{1}{N}\sum_{\mu=1}^K\ton*{ \xi_i^{\mu} \xi_j^{\mu} + \omega \tilde{\xi}_{ij}^{\mu}},
\end{equation}
where, again, $\tilde{\xi}_{ij}^{\mu}$ is a standard Gaussian random variable and $\omega$ is a real parameter that tunes the noise level.
In the RBM representation, this corresponds to a perfect learning, while defects emerge just in the associative network, see Fig.~\ref{fig:map_c}.
Notice that the coupling in (\ref{eq:noisy_couplings_c}) is linear in $\omega$ and it yields to relatively weak effects, in fact, we will show that in a low-load regime, $\omega$ can grow ``fast'' with the system size, without breaking retrieval capabilities.
		\begin{figure}
			\includegraphics[width=0.33\textwidth]{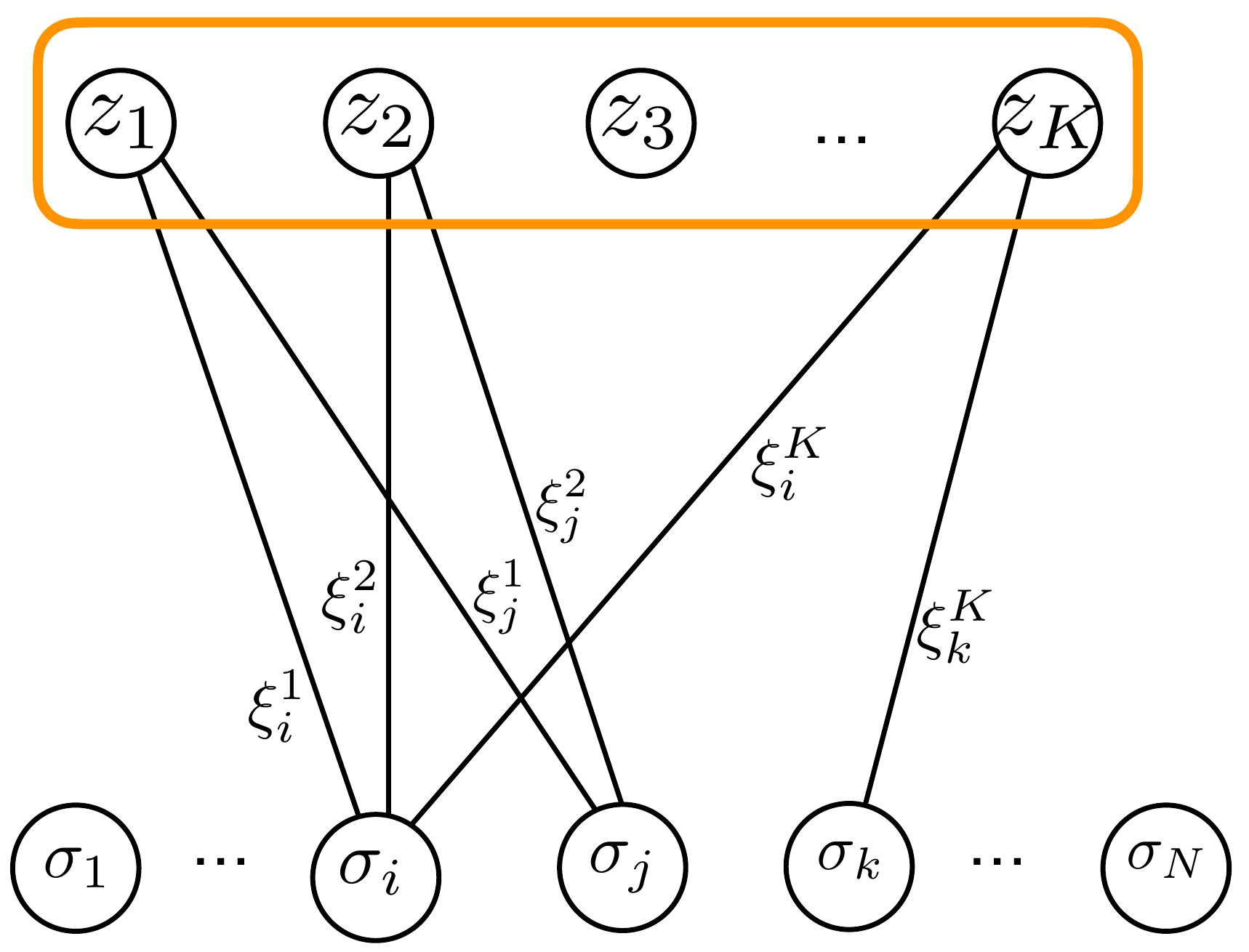}
			\caption{\textbf{RBM corresponding to shortcomings in the storage case.}  The machine is built over a hidden layer made of Gaussian neurons $\{z_{\mu}\}_{\mu=1,...,K}$ and a visible layer made of binary neurons $\{ \sigma_i \}_{i=1,...,N}$; in this case a neuron $z_{\mu}$ belonging to the hidden layer can interact with one neuron $\sigma_i$ belonging to the visible layer and the coupling is $\xi_i^{\mu}$, namely the patterns are correctly learnt and communications between the two layers is devoid of flaws. Since the machine is restricted, intra-layer interactions are not allowed. In the dual associative network the neurons interact pairwise ($p=2$) and the synaptic weight for the couplet ($\sigma_i, \sigma_j$) is $J_{ij} = \sum_{\mu}  \xi_i^{\mu} \xi_j^{\mu} + \omega \tilde{\xi}_{ij}^{\mu}$, as reported also in Eq.~\ref{eq:noisy_couplings_c}. This structure can be straightforwardly generalized for $p>2$. In this figure, again, seeking for clarity, only a few connections are drawn for illustrative purposes.}
\label{fig:map_c}
		\end{figure}
\newline
It is worth recalling that the problem of a HNN endowed with noisy couplings like in (\ref{eq:noisy_couplings_c}) has already been addressed in the past (see e.g., \cite{Amit,S-PRA1986,S-1987,TDC-PNAS1986,NTCD-EPL1986}).
In particular, Sompolinsky \cite{S-PRA1986,S-1987} showed that, in the high-load regime (i.e., $K \sim N$), the strength of noise affecting couplings still preserving retrieval is of order one. More precisely, denoting by $\delta_{ij}$ a centered Gaussian variable with variance $\delta^2$ and setting $J^s_{ij} = \sum_{\mu} \xi_i^{\mu} \xi_j^{\mu}/N + \delta_{ij}/\sqrt{N}$, he found that, as $\delta$ is fine tuned, the system capacity $\alpha$ is lowered and it vanishes for $\delta \approx 0.8$. From this result, one can conclude that the HNN is relatively robust to the presence of ``moderate levels'' of effective synaptic noise.
These findings are recovered in our investigations and suitably extended for $p>2$.
Notably, this kind of noise also includes, as a special example, the diluted network, where a finite fraction of the connections are cut randomly, still retaining a giant component \cite{Amit,S-PRA1986,S-1987,AABCT-JPA2013}. 
\end{enumerate}

Before concluding we need a few more definitions.
As aforementioned, we distinguish the tolerance with respect to interference among patterns (slow noise), which grows with $K$, and with respect to errors during learning or storing (synaptic noise), which grows with $\omega$. 
More quantitatively,  we set
\begin{eqnarray}
K = N^a, a \geq 0\\
\omega = N^b, b \geq 0,
\end{eqnarray}
and we introduce 
\begin{eqnarray}
\label{eq:alfa}
\alpha(b) := \max_{a ~ \textrm{s.t.} ~ \frac{S}{R} \lesssim 1} \frac{K}{N},\\
\label{eq:def_beta}
\beta(a) := \max_{b ~ \textrm{s.t.} ~  \frac{S}{R} \lesssim 1} \omega.
\end{eqnarray}
Finally, the Mattis magnetization, defined as 
\begin{equation}
m_{\mu} := \frac{1}{N} \sum_{i=1}^N \sigma_i \xi_i^{\mu}, ~~ \mu=1,...,K,
\end{equation}
is used to assess the retrieval of the $\mu$-th pattern.


	\section{The $p$-neuron Hopfield model with synaptic noise} \label{sec:DAM}

		The $p$-neuron Hopfield model is described by the Hamiltonian 
		\begin{equation}
			H^{(p)}(\vec \sigma, \boldsymbol \xi)=-\frac{1}{p!N^{p-1}}\sum_{\mu=1}^K\sum_{i_1,\ldots,i_p}\xi_{i_1}^{\mu}\ldots\xi_{i_p}^{\mu}\sigma_{i_1}\dots\sigma_{i_p},
			\label{eq:hamiltonian_monopartite_DAM}
		\end{equation}
		where the sum runs over all possible $p$-plets and self-interactions are excluded.
		This kind of model provides an example of dense AMs, which have been intensively studied in the last years (see e.g., \cite{Baldi,Bovier,AABCF-PRL2020,AABF-NN2020}).
		
		For even $p$, this model is thermodynamically equivalent to a RBM equipped with a hidden layer made of $K$ Gaussian neurons $\{ z_{\mu}\}_{\mu=1,...,K}$ and with a visible layer made of $N$ binary neurons $\{ \sigma_i \}_{i=1,...,N}$, but now couplings in the RBM are $(1+p/2)$-wise and include one hidden neuron and $p/2$ visible neurons, say $(z_{\mu}, \sigma_{i_1}, ..., \sigma_{i_{p/2}})$, and the related coupling in the $p$-neuron Hopfield model is $\xi_{i_1}^{\mu} ... \xi_{i_{p/2}}^{\mu}$.
     	To see the equivalence between this RBM and the model described by (\ref{eq:hamiltonian_monopartite_DAM}) we look at the RBM partition function and we perform the Gaussian integration to marginalize over the hidden units as
		\begin{eqnarray}
		\nonumber
		Z_{\textrm{RBM}}^{(p)}(\boldsymbol \xi) &=& \sum_{\vec \sigma} \prod_{\mu=1}^K \int d z_{\mu}  \frac{e^{- \frac{z_{\mu}^2}{2}}}{\sqrt{2\pi}}  e^{- \beta N^{\frac{1-p}{2}} (  \prod_{j=1}^{p/2}\sum_{i_j} \sigma_{i_j} \xi_{i_j}^{\mu}) z_{\mu}} \\
		&=& \sum_{\vec \sigma} \prod_{\mu=1}^K e^{- \frac{\beta'}{2} N^{1-p}  \prod_{j=1}^{p} \sum_{i_j} \sigma_{i_j} \xi_{i_j}^{\mu} },
		\end{eqnarray}
		where the inverse temperature $\beta$ has been properly rescaled into $\beta'$.

		Let us start the study of this system in the presence of slow noise only and let us check stability of the configuration $\vec \xi^1$, without loss of generality. By signal to noise analysis we write 
		\[
			h_i^{(p)}\xi_i^1=S+R^{(0)},
		\]
		where
		\begin{align*}
			&S^{}\sim1, \\
			&R^{(0)}=\frac{1}{p!N^{p-1}}\sum_{\mu=2}^K\sum_{i_2,\ldots,i_p}^N\xi_{i}^{\mu}\xi_{i_2}^{\mu}\ldots\xi_{i_p}^{\mu}.\xi^1_i\xi_{i_2}^1\dots\xi_{i_p}^1
		\end{align*}
		and, for large $N$, from the central limit theorem,
		\begin{equation}
			R^{(0)}\sim\frac{1}{N^{p-1}}\sqrt{KN^{p-1}}=\sqrt{\frac{K}{N^{p-1}}}.
			\label{eq:noise_DAM_interference_a}
		\end{equation}
		Recalling that the condition for retrieval is $R^{(0)}\lesssim S$, the highest load corresponds to  
	$K \sim N^{p-1}$, namely 
		\begin{equation}
		\alpha^{(p)} = N^{p-2},
		\end{equation}
		as previously proved in \cite{Baldi}.
			
		This result shows that increasing the number of interacting spins allows to arbitrary increase the tolerance versus slow noise. It is then natural to question if an analogous robustness can be obtained versus synaptic noise too. 
		In the next subsections we address this question for the three sources of noise outlined in Sec.~\ref{sec:defs}.


		\subsection{Noisy patterns} \label{ssec:noise_a}
			When the noise affects directly patterns constituting the dataset, using Eq.~\eqref{eq:noisy_patterns} we can write the product between the local field and a pattern, according to Eq.~\ref{eq:field}, as
			\begin{align*}
				h_i^{(p)}\xi_i^1&=\frac{1}{p!N^{p-1}}\sum_{\mu}^K\sum_{i_2,\ldots,i_p}^N\\
				&\ton*{\xi_{i}^{\mu}+\omega\tilde{\xi}_{i}^{\mu}}\ton*{\xi_{i_2}^{\mu}+\omega\tilde{\xi}_{i_2}^{\mu}}\ldots\ton*{\xi_{i_p}^{\mu}+\omega\tilde{\xi}_{i}^{\mu}}\xi_{i}^1\xi_{i_2}^1\ldots\xi_{i_p}^1.
			\end{align*}
			Splitting the sum into a signal $S$ and a noise $R$ term we obtain $h_i^{(p)}\xi_i^1=S+R$, with 
			\[
				S\sim1, \quad R=\sum_{n=0}^pR^{(n)}.
			\]	
			The quantity $R^{(0)}$ is the standard contribution due to slow noise given by Eq.~\eqref{eq:noise_DAM_interference_a}, while $\tilde{R}=\sum_{n=1}^pR^{(n)}$ derives from the presence of synaptic noise. To simplify the following formulas we rename $i$ as $i_1$ and write this last contribution as 
			\begin{align*}
				\tilde{R}=&\frac{1}{p!N^{p-1}}\sum_{\mu}^K\sum_{i_2,\ldots,i_p}^N\xi_{i_1}^1\xi_{i_2}^1\ldots\xi_{i_p}^1\\
				&\left\{\underbrace{\omega\sum_{(i_x)}\xi_{i_1}^{\mu}\ldots\tilde{\xi}_{i_x}^{\mu}\ldots\xi_{i_p}^{\mu}}_{R^{(1)}}+\right.\\
				&\hspace{1em}+\underbrace{\omega^2\sum_{(i_x, i_y)}\xi_{i_1}^{\mu}\ldots\tilde{\xi}_{i_x}^{\mu}\ldots\tilde{\xi}_{i_y}^{\mu}\ldots\xi_{i_p}^{\mu}}_{R^{(2)}}+\\
				&\hspace{1em}+\underbrace{\omega^3\sum_{(i_x, i_y, i_z)}\xi_{i_1}^{\mu}\ldots\tilde{\xi}_{i_x}^{\mu}\ldots\tilde{\xi}_{i_y}^{\mu}\ldots\tilde{\xi}_{i_z}^{\mu}\ldots\xi_{i_p}^{\mu}}_{R^{(3)}}+\\
				&\hspace{1.7em}\vdots\\
				&\left.\hspace{1em}+\ \underbrace{\omega^p\tilde{\xi}_{i_1}^{\mu}\tilde{\xi}_{i_2}^{\mu}\tilde{\xi}_{i_3}^{\mu}\ldots\tilde{\xi}_{i_p}^{\mu}}_{R^{(p)}}\right\},
			\end{align*}
			where $\sum_{(i_{a_1}\ldots i_{a_n})}$ denotes the sum over all possible permutations of $n$ indices chosen from $i_1\ldots i_p$.
			Using the central limit theorem (as explained in details for $p=4$ in the Appendix \ref{sec:4-DAM}) we obtain that
			\begin{align*} 
				R^{(n)}\sim\frac{\omega^n}{N^{p-1}}&\left[N^{p-n}\ton*{N^{1/2}}^{n-1}+N^{p-(n-1)}\ton*{N^{1/2}}^n\right.\\
				&\left.+\sqrt{KN^{p-1}}\right].
			\end{align*}
			Then, at leading order, it holds 
			\[
				\tilde{R}\sim\frac{1}{N^{p-1}}\qua*{\ton*{\sum_{n=1}^p\omega^nN^{p-n}\ton*{N^{1/2}}^{n-1}}+\omega^p\sqrt{KN^{p-1}}}.
			\]
			Therefore, overall, the noise $R=R^{(0)}+\tilde{R}$ scales as  
			\[
				R\sim\qua*{\sum_{n=1}^p\omega^nN^{1-n}\ton*{N^{1/2}}^{n-1}}+\omega^p\sqrt{\frac{K}{N^{p-1}}}.
			\]
			Recalling that $S\sim1$, we conclude that retrieval is possible provided that $\omega\sim1$, independently of the number $K$ of stored patterns (up to $K\sim N^{p-1}$). This implies that a diverging synaptic noise (i.e., $\omega \sim \mathcal O (N^b), b>0$) can not be handled by the system even if the number $p$ of spins interacting and, accordingly, the number of links, is arbitrarly increased.	
			
\begin{figure}
			\includegraphics[width=0.45\textwidth]{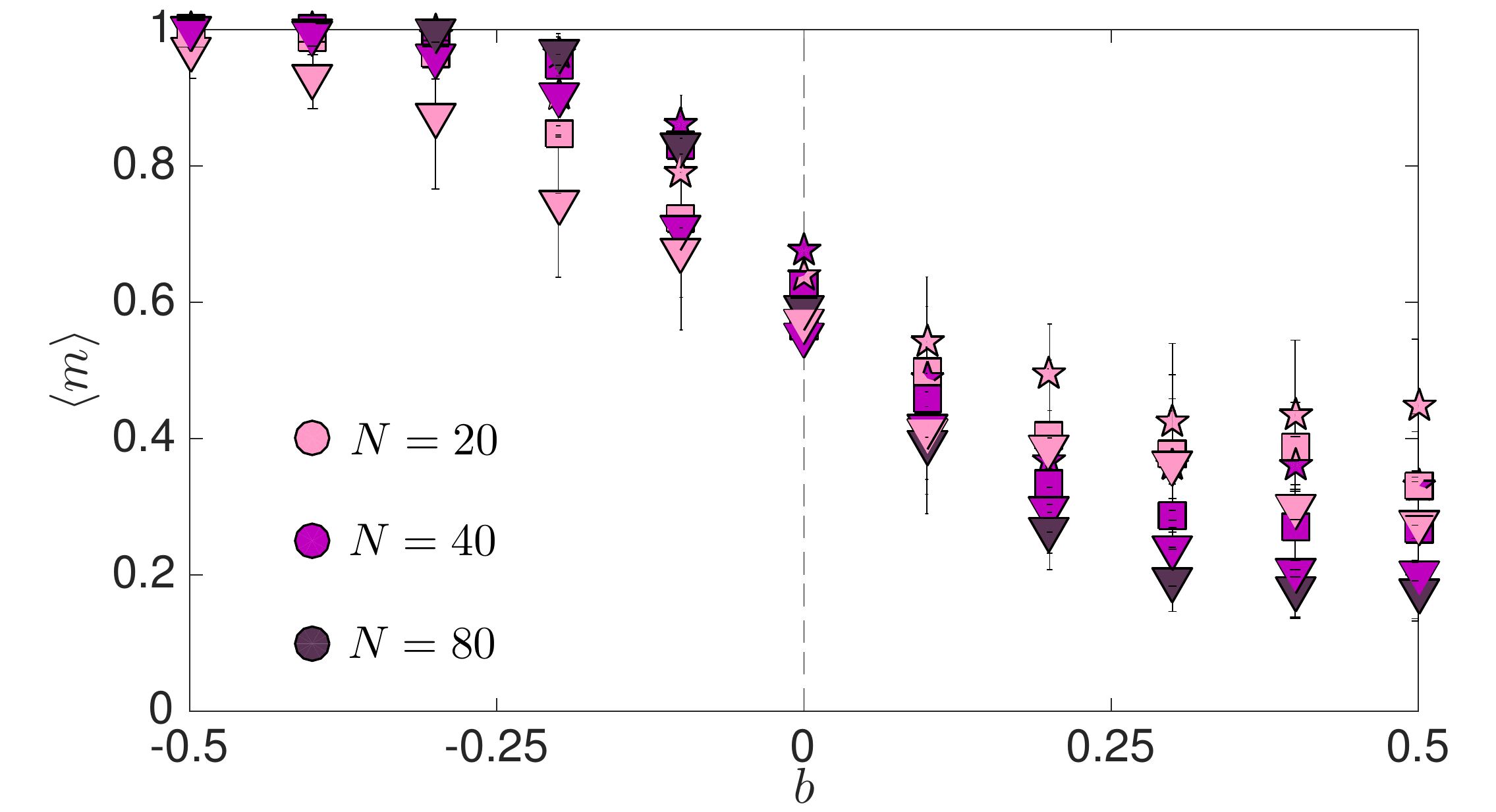}
			\caption{\textbf{Numerical simulations for the $p$-neuron Hopfield model endowed with noisy patterns ($p>2$).} We simulated the evolution of a $p$-neuron Hopfield model, with $p=3$ ($\triangle$), $p=4$ ($\square$), and $p=5$ ($\star$), under the dynamics (\ref{eq:dynamic}) and using as starting state $\boldsymbol \sigma = \boldsymbol {\xi^{\mu}}$, finally collecting the Mattis magnetizations $m_{\mu}$, for $\mu=1,...,K$. Here we set $K=N$ and $\omega=N^b$, where $b$ is varied in $[-0.5, 0.5]$, and we plot the average magnetization $\langle m \rangle$ versus $b$; the magnetization is averaged over $\mu$ and over $M=10$ realizations of the patterns $\boldsymbol \eta$, as defined in (\ref{eq:noisy_patterns}), the standard deviation is represented by the errorbar. Three different sizes are considered $N=20$, $N=40$, $N=80$, as explained by the legend. The vertical dashed line is set at $b=0$ and highlights the threshold for retrieval, as stated in the main text.}
\label{fig:rumore_a}
		\end{figure}

\begin{figure}
			\includegraphics[width=0.45\textwidth]{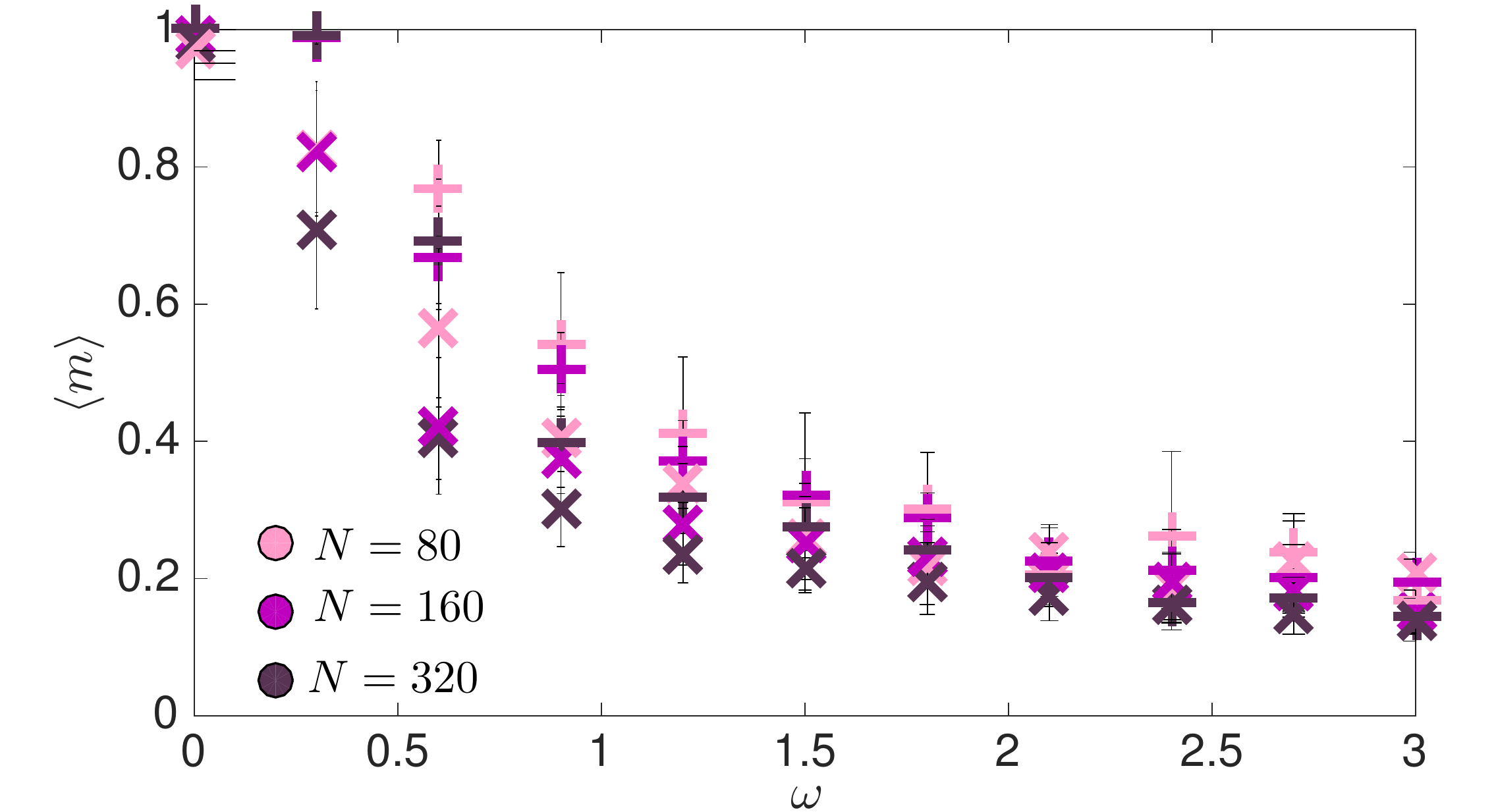}
			\caption{\textbf{Numerical simulations for the Hopfield model with pairwise couplings ($p=2$) endowed with noisy patterns.} 
			We run numerical simulation as explained in the caption of Fig.~\ref{fig:rumore_a} but setting $p=2$ and varying $\omega$ linearly in $[0,3]$. We compare two loads: $K/N = 0.125$ ($\times$) and $K/N=0.04$ ($+$). Notice that, in both case, even small values of $\omega$ yield to a breakdown of retrieval.}
\label{fig:rumore_a_p2}
		\end{figure}

			This result is checked numerically as shown in Fig.~\ref{fig:rumore_a}. In particular, we notice that, as long as $\omega$ remains finite (or vanishing) while the size $N$ is increased, i.e., as long as $b \leq 0$, the Mattis magnetization corresponding to the input pattern is non null and the system can retrieve. The transition between a retrieval and a non-retrieval regime is sharper when the network size is larger. 
		In Fig.~\ref{fig:rumore_a_p2} we focus on $p=2$ and we set the ratio $K/N < \alpha(b=0) \approx 0.138$, while we perform a fine tuning by varying $\omega \in [0,3]$. As expected, even small values of $\omega$ are sufficient to break down retrieval capabilities.

		\subsection{Noisy learning}
\begin{figure}
			\includegraphics[width=0.45\textwidth]{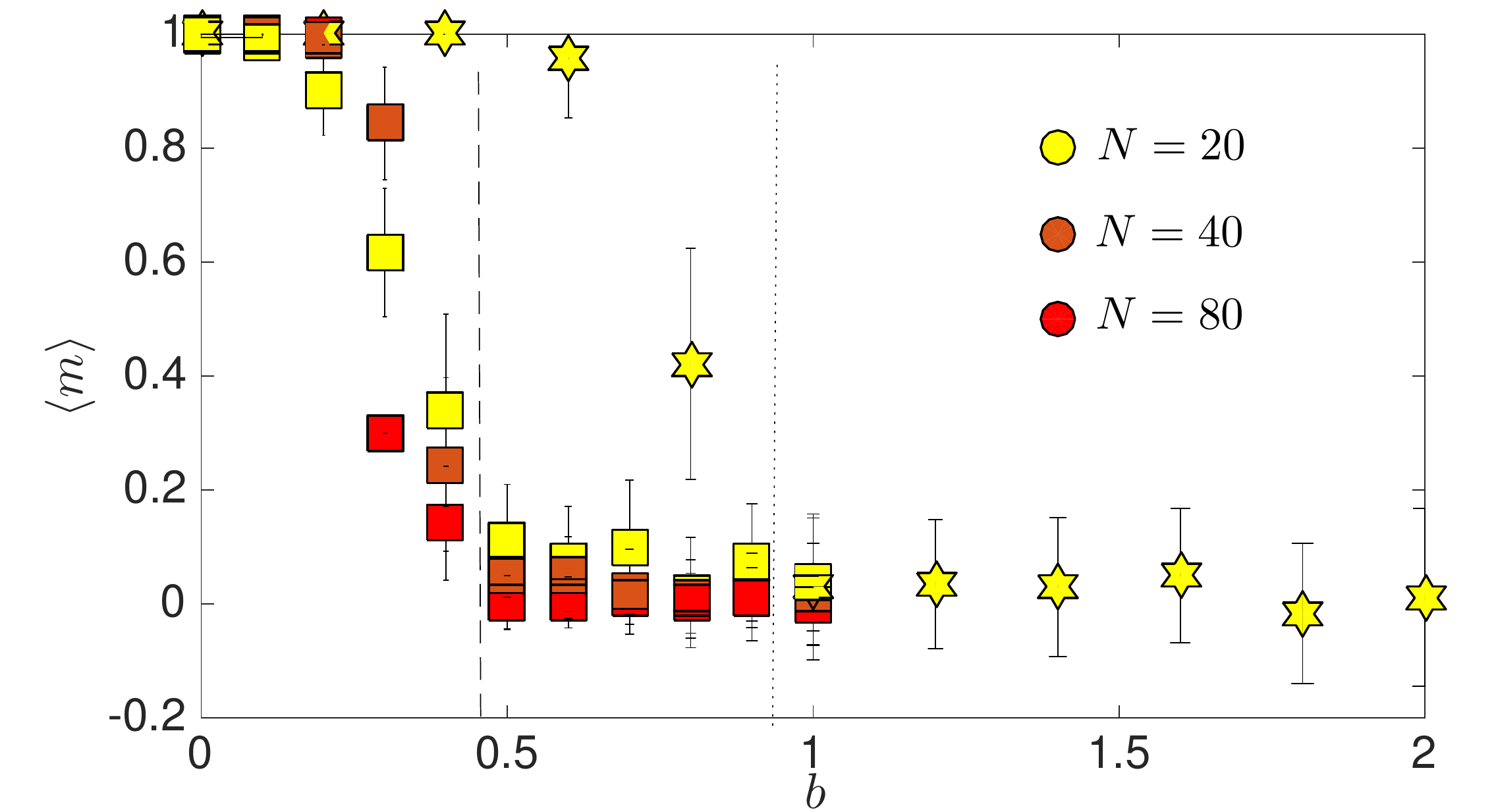}
			\caption{\textbf{Numerical simulations for the $p$-neuron Hopfield affected by noisy learning ($p>2$).} We simulated the evolution of a $p$-neuron Hopfield model, with $p=4$ ($\square$) and $p=6$ ($\ast$), under the dynamics (\ref{eq:dynamic}) and using as starting state $\boldsymbol \sigma = \boldsymbol {\xi^{\mu}}$, finally collecting the Mattis magnetizations $m_{\mu}$, for $\mu=1,...,K$. Here we set $K=N$ and $\omega=N^b$, where $b$ is varied in, respectively, $[0, 1]$ and in $[0,2]$, and we plot the average magnetization $\langle m \rangle$ versus $b$; the magnetization is averaged over $\mu$ and over $M=10$ realizations of the patterns $\boldsymbol \eta$, as defined in (\ref{eq:noisy_patterns}), the standard deviation is rapresented by the errorbar. Three different sizes are considered $N=20$, $N=40$, $N=80$, as explained by the legend. The dashed and dotted vertical lines are set at $b=0.5$ and $b=1.0$, which represent the thresholds for retrieival for, respectively, $p=4$ and $p=6$, according to (\ref{eq:agree}).}
\label{fig:rumore_b}
		\end{figure}

		Let us now consider the AM corresponding to imperfect learning as depicted in Fig.~\ref{fig:map_b}. 
		This equals to say that the noise affects the $(p/2+1)$-component tensor 
			\[
				\eta_{i_1\ldots i_{p/2}}^{\mu}=\xi_{i_1}^{\mu}\ldots\xi_{i_{p/2}}^{\mu}+\omega\tilde{\xi}^{\mu}_{i_1\ldots i_{p/2}},
			\]
			in such a way that the coupling between neurons is
						\begin{align} \label{eq:coupling_b}
			J_{i_1, ..., i_p} = \sum_{\mu} &(\xi_{i_1}^{\mu}\ldots\xi_{i_{p/2}}^{\mu}+\omega\tilde{\xi}^{\mu}_{i_1\ldots i_{p/2}})\nonumber \\
			& \times (\xi_{i_{1+p/2}}^{\mu}\ldots\xi_{i_{p}}^{\mu}+\omega\tilde{\xi}^{\mu}_{i_{1+p/2}\ldots i_{p}})
			\end{align}
			Notice that this picture is possible only for even $p$ and constitutes a generalization of the system studied in \cite{ACF-JPA2020}. The product between the local field and the pattern $\boldsymbol{\xi^1}$ candidate for retrieval reads
			\begin{align*}
				\xi_i^1h_i=\frac{1}{p!N^{p-1}}&\sum_{\mu}^K\sum_{i_2,\ldots,i_p}^N\ton*{\xi_{i_1}^{\mu}\ldots\xi_{i_{p/2}}^{\mu}+\omega\tilde{\xi}^{\mu}_{i_1\ldots i_{p/2}}}\\
				&\ton*{\xi_{i_{p/2+1}}^{\mu}\ldots\xi_{i_{p}}^{\mu}+\omega\tilde{\xi}^{\mu}_{i_{p/2+1}\ldots i_{p}}}\xi_{i}^1\xi_{i_2}^1\ldots\xi_{i_p}^1.
			\end{align*}
			Again, we can split this quantity into a signal $S$ and noise $R=\sum_{n=0}^2R^{(n)}$ term, the signal and the zeroth order of noise are, as already shown,
			\[
				S\sim1, \quad R^{(0)}\sim\sqrt{\frac{K}{N^{p-1}}}.
			\]
			The first order contribution is 
			\begin{align*}
				R^{(1)}=\frac{\omega}{p!N^{p-1}}&\sum_{\mu=1}^K\sum_{i_2,\ldots,i_p}^N\left(\xi_{i_1}^{\mu}\ldots\xi_{i_{p/2}}^{\mu}\tilde{\xi}^{\mu}_{i_{p/2+1}\ldots i_{p}}\right.+\\
				&\left.\xi_{i_{p/2+1}}^{\mu}\ldots\xi_{i_{p}}^{\mu}\tilde{\xi}^{\mu}_{i_1\ldots i_{p/2}}\right)\xi_{i}^1\xi_{i_2}^1\ldots\xi_{i_p}^1,
			\end{align*}
			and, in the limit of large network size (for more details we refer to Appendix \ref{sec:4-DAM} were calculations for $p=4$ are reported), 
			\begin{align*}
				R^{(1)}\sim\frac{\omega}{N^{p-1}}&\left[N^{p/2}\ton*{N^{1/2}}^{p/2-1}+N^{p/2-1}\ton*{N^{1/2}}^{p/2}\right.\\
				&\left.+\sqrt{KN^{p-1}}\right].
			\end{align*}
			Similarly, the second order contribution is of the form 
			\begin{align*}
				R^{(2)}&=\frac{\omega^2}{p!N^{p-1}}\sum_{\mu=1}^K\sum_{i_2,\ldots,i_p}^N\tilde{\xi}^{\mu}_{i_1\ldots i_{p/2}}\tilde{\xi}^{\mu}_{i_{p/2+1}\ldots i_p}\xi_{i}^1\xi_{i_2}^1\ldots\xi_{i_p}^1\sim\\
				&\sim\frac{\omega^2}{N^{p-1}}\sqrt{KN^{p-1}}.
			\end{align*}
			We then deduce that the noise $R$ scales as 
			\begin{align*}
				R\sim\frac{1}{N^{p-1}}&\left \{ \omega\qua*{N^{p/2}\ton*{N^{1/2}}^{p/2-1}+N^{p/2-1}\ton*{N^{1/2}}^{p/2}}+\right.\\
				&\left.+\sqrt{KN^{p-1}}\ton*{1+\omega+\omega^2}\right \},
			\end{align*}
			and therefore, neglecting subleading contributions, we can write 
			\[
				R\sim\omega N^{1/2-p/4}+\omega^2\sqrt{K}N^{1/2-p/2}.
			\]
			Setting $K\sim N^a$ and $\omega\sim N^b$ the condition for retrieval reads 
			\[
				N^{1/2-p/4+b}+N^{(1-p+a)/2+2b}\lesssim 1.
			\]	
			By comparing the scaling of the two terms in the r.h.s. of the previous equation we see that the former diverges with $N$ if $b>p/4-1/2$, while the latter diverges if $b>p/4-(1+a)/4$. This implies that when $a\leq1$ the first term dominates the signal-to-noise analysis and the extremal condition for retrieval reads $b=(p-2)/4$.
			Therefore, the tolerance versus synaptic noise is 
			\[
				\beta_p(a)\sim N^{p/4-1/2} \ \text{for}\ a\leq1.
			\]
			Conversely, if $a>1$, the second term prevails and consequently the extremal condition for retrieval becomes $b=(p-1-a)/4$, and the tolerance is
			\[
				\beta_p(a)\sim N^{p/4-(1+a)/4} \ \text{for}\ 1<a<p-1.
			\]
			Note that in this case the tolerance depends on $a$, that is, on the network load. This shows that storing and tolerance are intimately tangled: the larger the load and the smaller the noise that can be handled.
			In particular, at low load, so for $a=1$, the tolerance reads 
			\begin{equation} \label{eq:agree}
				\beta_p(1)\sim N^{p/4-1/2},
			\end{equation}
			as corroborated numerically in Fig.~\ref{fig:rumore_b}.
			For $p=2$ this kind of noise reduces to the case discussed in Subsec.~\ref{ssec:noise_a} and consistenly we get $\beta_2(1)\sim 1$.
			
		\subsection{Noisy storing}
\begin{figure}
			\includegraphics[width=0.45\textwidth]{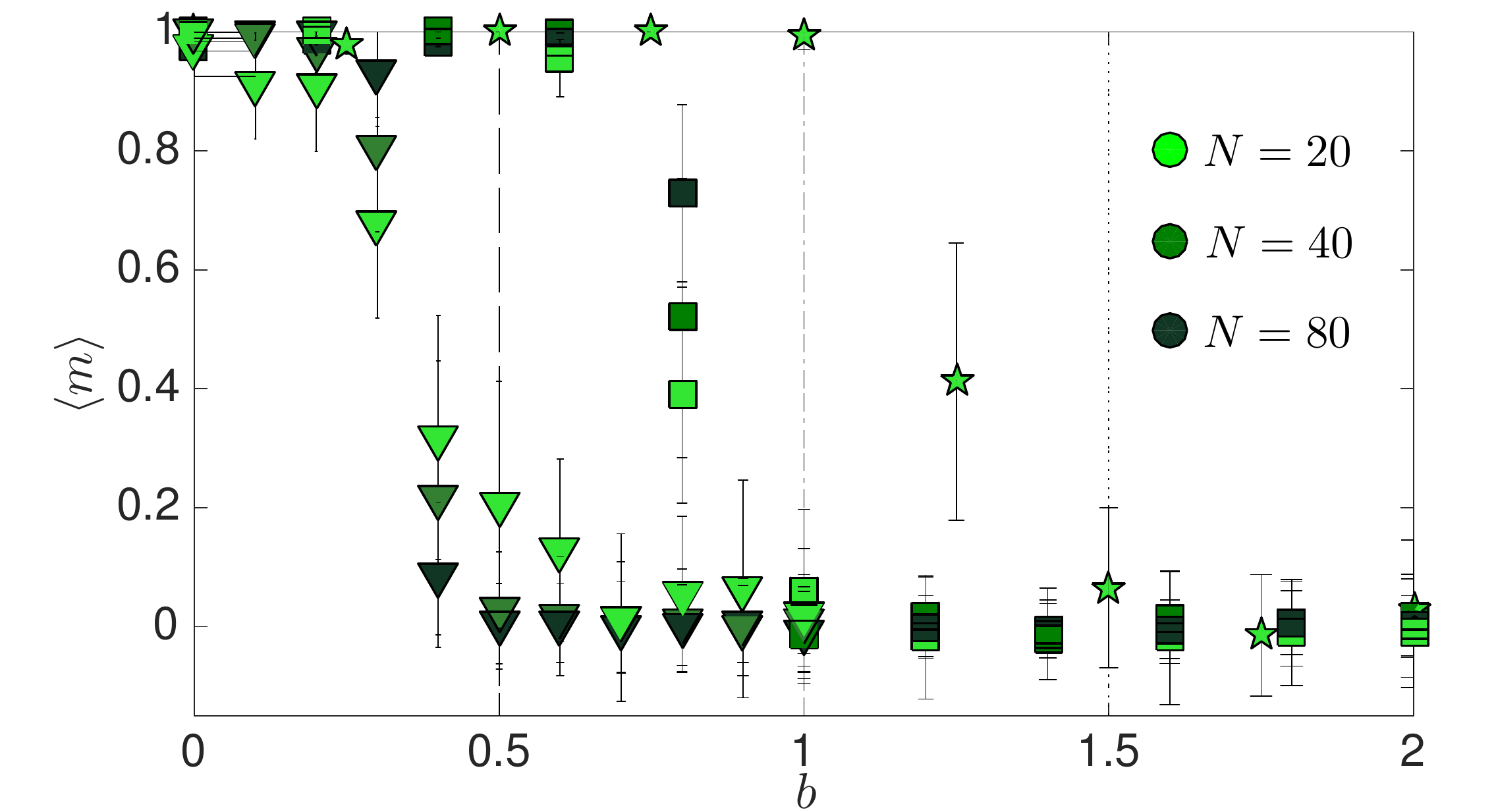}
			\caption{\textbf{Numerical simulations for the $p$-neuron Hopfield affected by noisy storing ($p>2$).} We simulated the evolution of a $p$-neuron Hopfield model, with $p=3$ ($\triangle$), $p=4$ ($\square$), and $p=5$ ($\star$), under the dynamics (\ref{eq:dynamic}) and using as starting state $\boldsymbol \sigma = \boldsymbol {\xi^{\mu}}$, finally collecting the Mattis magnetizations $m_{\mu}$, for $\mu=1,...,K$. Here we set $K=N$ and $\omega=N^b$, where $b$ is varied in $[0, 2]$, and we plot the average magnetization $\langle m \rangle$ versus $b$; the magnetization is averaged over $\mu$ and over $M=10$ realizations of the couplings $\boldsymbol J$, as defined in (\ref{eq:noisy_couplings_c}), the standard deviation is rapresented by the errorbar. Three different sizes are considered $N=20$, $N=40$, $N=80$, as explained by the legend.}
\label{fig:rumore_c}
		\end{figure}

\begin{figure}
			\includegraphics[width=0.45\textwidth]{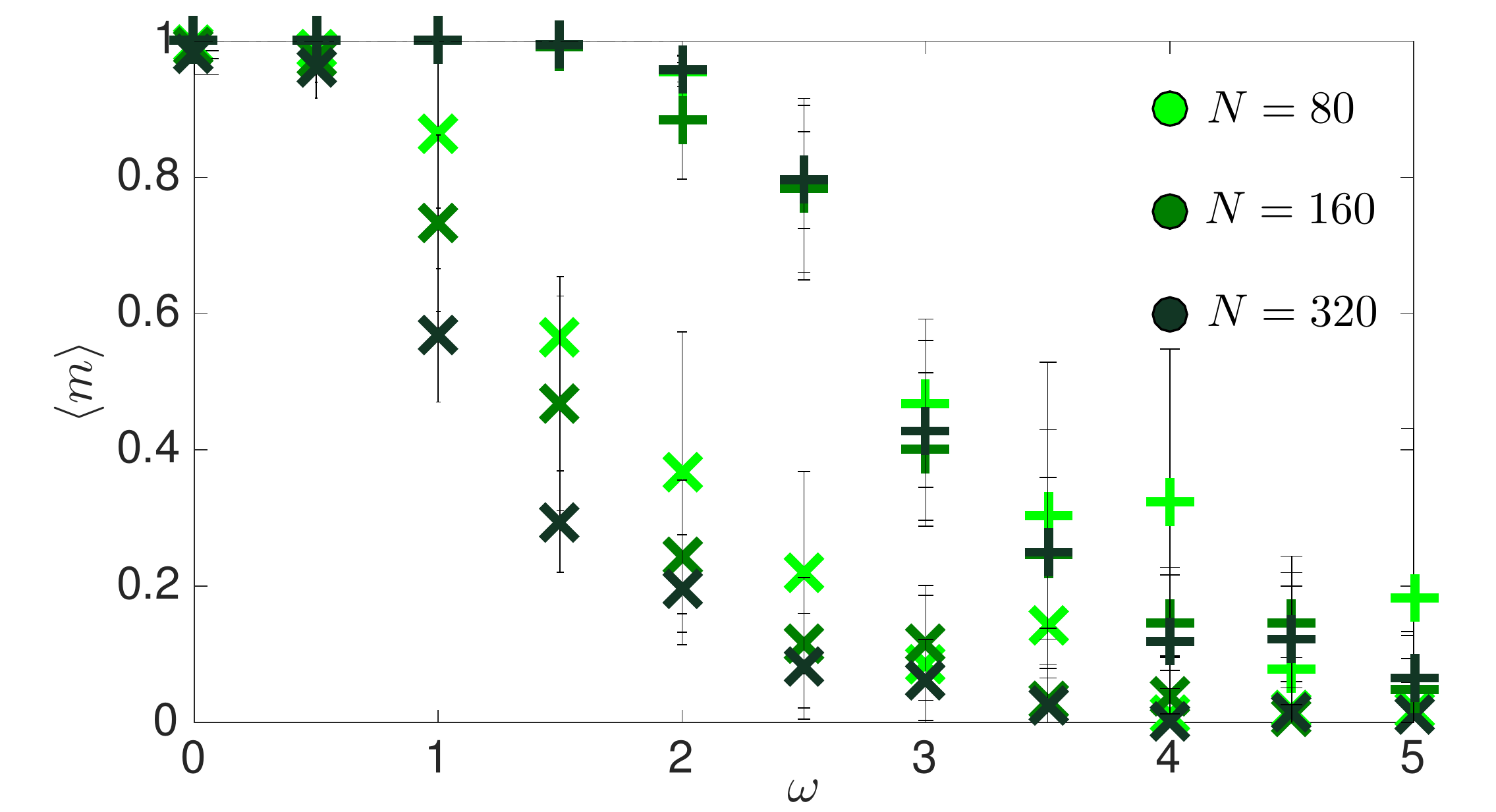}
			\caption{\textbf{Numerical simulations for the Hopfield model with pairwise couplings ($p=2$) endowed with noisy couplings.} 
			We run numerical simulation as explained in the caption of Fig.~\ref{fig:rumore_c} but setting $p=2$ and varying $\omega$ linearly in $[0,5]$. We compare two loads: $K/N = 0.125$ ($\times$) and $K/N=0.04$ ($+$). Notice that, in both case, as $\omega$ is relatively large the retrieval is lost.}
\label{fig:rumore_c_p2}
		\end{figure}

			Finally, we consider noise acting directly on couplings, 
			\begin{equation}
			J_{i_1\ldots i_p}^{\mu} = \sum_{\mu} \eta_{i_1\ldots i_p}^{\mu},
			\end{equation}
			where $\eta_{i_1\ldots i_p}^{\mu}$ is the $(p+1)$-component tensor
			\[
				\eta_{i_1\ldots i_p}^{\mu}=\xi_{i_1}^{\mu}\ldots\xi_{i_p}^{\mu}+\omega\tilde{\xi}_{i_1\ldots i_p}^{\mu}.
			\]
			Still following the prescription coded by Eq.~\ref{eq:field}, the product between the local field $h_i$ and $\xi_i^1$ is 
			\[
				h_i\xi_i^1=\frac{1}{p!N^{p-1}}\sum_{\mu=1}^K\sum_{i_2\ldots i_p}^N\ton*{\xi_{i_1}^{\mu}\ldots\xi_{i_p}^{\mu}+\omega\tilde{\xi}_{i_1\ldots i_p}^{\mu}}\xi_{i_1}^1\ldots\xi_{i_p}^1
			\]
			The signal scales as $S\sim1$, while the noise is composed on solely two contributions: zeroth and first order. We have already computed the former
			\[
				R^{(0)}\sim\sqrt{\frac{K}{N^{p-1}}},
			\]
			and, as for the latter, it holds 
			\[
				R^{(1)}=\frac{\omega}{p!N^{p-1}}\sum_{\mu=1}^K\sum_{i_2\ldots i_p}^N\tilde{\xi}_{i_1\ldots i_p}^{\mu}\xi_{i_1}^1\ldots\xi_{i_p}^1\sim\omega\sqrt{\frac{K}{N^{p-1}}}.
			\]
			Therefore, 
			\[
				R=R^{(0)}+R^{(1)}\sim\sqrt{\frac{K}{N^{p-1}}}\ton*{1+\omega}\sim\omega\sqrt{\frac{K}{N^{p-1}}}.
			\]
			Setting, as before, $K\sim N^a$ and $\omega\sim N^b$ the condition for retrieval becomes
			\[
				N^{(a-p+1)/2+b}\sim1\to b=\frac{p-1-a}{2}.
			\]
			This implies that the tolerance versus pattern noise is
			\begin{equation}
				\beta_p(a)\sim N^{(p-1-a)/2}\ \text{for}\ a\leq p-1.
				\label{eq:tolerance_C}
			\end{equation}
			This is succesfully checked numerically in Fig.~\ref{fig:rumore_c}.
			The particular case $p=2$ is considered in Fig.~\ref{fig:rumore_c_p2}.
			Again, as pointed out in the previous section, tolerance and load are sides of the same coin: an increase of the latter results in a decrease of the former.

			A similar problem, for the $p=2$ Hopfield model, has been studied by Sompolinsky \cite{S-1987, S-PRA1986}. In particular the following couplings have been considered
			\[
				J^s_{ij}=\ton*{\frac{1}{N}\sum_{\mu=1}^K\xi_i^{\mu}\xi_j^{\mu}}+\underbrace{\frac{\delta_{ij}}{\sqrt{N}}}_{\tilde{J}_{ij}^s}.
			\]
			Here $\delta_{ij}$ are Gaussian variables with null mean and variance $\delta^2$, while $\tilde{J}_{ij}^s$ represents the correction to Hebbian couplings due to noise. Focusing on the high load regime, that is $K\sim N$, retrieval was found to be possible provided that $\delta\lesssim0.8$.
			We can easily map noise defined by Eq.~\eqref{eq:noisy_couplings_c} into this notation, indeed
			\[
				J_{ij} = \frac{1}{N}\sum_{\mu=1}^K\ton*{\xi_i^{\mu} \xi_j^{\mu} + \omega \tilde{\xi}_{ij}^{\mu}}=\frac{1}{N}\sum_{\mu=1}^K\xi_i^{\mu} \xi_j^{\mu}+\underbrace{\frac{\omega}{N}\sum_{\mu=1}^K\tilde{\xi}_{ij}^{\mu}}_{\tilde{J}_{ij}}.
			\]
			As a consequence, in our framework the noisy contribution to couplings reads
			\[
				\tilde{J}_{ij}=\frac{\omega}{N}\sum_{\mu=1}^K\tilde{\xi}_{ij}^{\mu}=\frac{\omega_{ij}\sqrt{K}}{N},
			\]		
			where $\omega_{ij}$ are Gaussian variables with null mean and variance $\omega^2$. Considering the high load regime we then obtain
			\[
				\tilde{J}_{ij}\sim\frac{\omega_{ij}\sqrt{N}}{N}=\frac{\omega_{ij}}{\sqrt{N}}.
			\]
			This shows that $\omega_{ij}$ is the counterpart of $\delta_{ij}$ and, therefore, that $\omega$ plays the same role of $\delta$. Recalling Eq.~\eqref{eq:tolerance_C} and setting $p=2$ and $a=1$ we conclude that retrieval is possible provided that $\omega\lesssim1$. This result is in perfect agreement with Sompolinsky's bound $\delta\lesssim0.8$ and also with the simulations we run.
			
		\section{Conclusions} \label{sec:conclusions}
		In this work we considered dense AMs and we investigated the role of density in preventing retrieval break-down due to noise. In particular, we allow for noise stemming from pattern interference (i.e., slow noise) and for noise stemming from uncertainties during learning or storing (i.e., synaptic noise), while fast noise is neglected. 
		Synaptic noise ultimately affects the synaptic couplings among neurons making up the network and we envisage different ways to model it, mimicking different physical situations. In fact, since couplings encode for the pieces of information previously learned, we can account for the following scenarios: $i$. information during learning is provided corrupted, $ii.$ information is supplied correctly but is imperfectly learned, $iii.$ information is well supplied and learned but storing is not accurate.
		These cases are discussed leveraging on the duality between AM and RBMs  \cite{AABCF-PRL2020, Bernacchia, ABGGM-PRL2020,ABDG-NN2013,AGST-PRE2017,AGST-PRE2018}.
		\newline
		Investigations were led analytically (via signal-to-noise approach) and numerically (via Monte Carlo simulations) finding that, according to the way synaptic noise is implemented, effects on retrieval can vary qualitatively.
		As long as the dataset is provided correctly during learning, synaptic noise can be annihilated by increasing redundancy (i.e., by letting neurons interact in relatively large cliques or work in a low-load regime). On the other hand, if, during learning, the machine was presented to corrupted pieces of information, it will learn noise as well and the correct information can be retrieved only if the original corruption is non diverging, no matter how redundant the network is.

		\section*{Acknowledgments}
EA is grateful 
to Universit\`a Sapienza di Roma (Progetto Ateneo RG11715C7CC31E3D) for financial support.


\appendix

\section{The $4$-neuron Hopfield model} \label{sec:4-DAM}		
In this appendix we set $p=4$ and we go through signal-to-noise calculations in detail.

		The $4$-neuron Hopfield model is described by the Hamiltonian 
		\begin{equation}
			H^{(4)}(\vec \sigma)=-\frac{1}{4!N^{3}}\sum_{\mu=1}^K\sum_{i,j,k,l}^{N}\xi_{i}^{\mu}\xi_{j}^{\mu}\xi_{k}^{\mu}\xi_{l}^{\mu}\sigma_{i}\sigma_{j}\sigma_{k}\sigma_{l}.
			\label{eq:hamiltonian_monopartite_4-DAM}
		\end{equation}
		where the sum is meant without self-interaction.
		Let us start the study of this system in presence of slow noise only and let us check stability of the configuration $\vec \xi^1$, without loss of generality. By signal to noise analysis we write 
		\[
			h_i\xi_i^1=S+R^{(0)},
		\]
		with
		\begin{align*}
			&S=\frac{1}{N^{3}}\sum_{j,k,l}\xi_{i}^{1}\xi_{j}^{1}\xi_{k}^{1}\xi_{l}^{1}\xi_{i}^{1}\xi_{j}^{1}\xi_{k}^{1}\xi_{l}^{1}\sim1,\\
			&R^{(0)}=\frac{1}{N^{3}}\sum_{\mu=2}^K\sum_{jkl}^N\xi_{i}^{\mu}\xi_{j}^{\mu}\xi_{k}^{\mu}\xi_{l}^{\mu}\xi^1_i\xi_k^1\xi^1_l \sim\frac{\sqrt{KN^3}}{N^3}=\sqrt{\frac{K}{N^3}},
		\end{align*}
		where asymptotic expressions are obtained exploiting the central limit theorem.
		Recalling that the condition for retrieval is $R^{(0)}\lesssim S$, the highest load corresponds to  
	$K \sim N^3$, namely 
		\begin{equation}
		\alpha^{(4)} = N^{2}.
		\end{equation}

		\subsection{Noisy patterns}	
			We now turn to the case in which the network is affected by pattern noise. We begin considering a situation in which the noise arises directly from patterns, in particular we suppose that the network stores the following vectors
			\begin{equation}
				\eta_i^{\mu}=\xi_i^{\mu}+\omega\tilde{\xi}_i^{\mu},
				\label{eq:noisy_patterns_app}
			\end{equation}		
			where $\xi_i^{\mu}$ are the patterns we would like to memorize, while $\tilde{\xi}_i^{\mu}$ are i.i.d. Gaussian variables with null mean and unitary variance. In order to study the stability of $\xi_i^1$ we consider the local field acting on it 
			\begin{align*}
				h_i\xi_i^1=\frac{1}{N^3}\sum_{\mu=1}^K\sum_{j,k,l}^N&\ton*{\xi_i^{\mu}+\omega\tilde{\xi}_{i}^{\mu}}\ton*{\xi_j^{\mu}+\omega\tilde{\xi}_{j}^{\mu}}\\
				&\ton*{\xi_k^{\mu}+\omega\tilde{\xi}_{k}^{\mu}}\ton*{\xi_l^{\mu}+\omega\tilde{\xi}_{l}^{\mu}}\xi_i^1\xi_j^1\xi_k^1\xi_l^1.
			\end{align*}
			We split this sum in signal $S$ ad noise $R=\sum_{i=0}^4R^{(i)}$. The signal and the zeroth order of noise are straightforward 
			\[
				S\sim1,
			\]
			\[
				R^{(0)}\sim\sqrt{\frac{K}{N^3}}.
			\]
			The first order is 
			\begin{align*}
				R^{(1)}=\frac{\omega}{N^3}\sum_{\mu=1}^K\sum_{j,k,l}^N\xi_i^1\xi_j^1\xi_k^1\xi_l^1&\left(\tilde{\xi}_i^{\mu}\xi_j^{\mu}\xi_k^{\mu}\xi_l^{\mu}+\xi_i^{\mu}\tilde{\xi}_j^{\mu}\xi_k^{\mu}\xi_l^{\mu}\right.\\
				&\left.\xi_i^{\mu}\xi_j^{\mu}\tilde{\xi}_k^{\mu}\xi_l^{\mu}+\tilde{\xi}_i^{\mu}\xi_j^{\mu}\xi_k^{\mu}\xi_l^{\mu}\right).
			\end{align*}
			That is 
			\[
				R^{(1)}=\frac{\omega}{2N^3}\sum_{\mu=1}^K\sum_{j,k,l}^N\xi_i^1\xi_j^1\xi_k^1\xi_l^1\ton*{\underbrace{\tilde{\xi}_i^{\mu}\xi_j^{\mu}\xi_k^{\mu}\xi_l^{\mu}}_{(a)}+3\underbrace{\xi_i^{\mu}\tilde{\xi}_j^{\mu}\xi_k^{\mu}\xi_l^{\mu}}_{(b)}}.
			\]
			Let us study the two terms separately 
			\begin{align*}
				(a)=&\frac{\omega}{N^3}\sum_{j,k,l}^N\xi_i^1\xi_j^1\xi_k^1\xi_l^1\tilde{\xi}_i^{1}\xi_j^{1}\xi_k^{1}\xi_l^{1}+\\
				&+\frac{\omega}{N^3}\sum_{\mu=2}^K\sum_{j,k,l}^N\xi_i^1\xi_j^1\xi_k^1\xi_l^1\tilde{\xi}_i^{\mu}\xi_j^{\mu}\xi_k^{\mu}\xi_l^{\mu}=\\
				&=\frac{\omega}{N^3}\ton*{\sum_{j,k,l}^N\xi_i^1\tilde{\xi}_i^{1}+\sum_{\mu=2}^K\sum_{j,k,l}^N\xi_i^1\xi_j^1\xi_k^1\xi_l^1\tilde{\xi}_i^{\mu}\xi_j^{\mu}\xi_k^{\mu}\xi_l^{\mu}},
			\end{align*}
			it then follows
			\[
				(a)\sim\omega\ton*{1+\sqrt{\frac{K}{N^3}}}.
			\]
			For what concerns the other term
			\begin{align*}
				(b)=&\frac{\omega}{N^3}\sum_{j,k,l}^N\xi_i^1\xi_j^1\xi_k^1\xi_l^1\xi_i^{1}\tilde{\xi}_j^{1}\xi_k^{1}\xi_l^{1}+\\
				&+\frac{\omega}{N^3}\sum_{\mu=2}^K\sum_{j,k,l}^N\xi_i^1\xi_j^1\xi_k^1\xi_l^1\xi_i^{\mu}\tilde{\xi}_j^{\mu}\xi_k^{\mu}\xi_l^{\mu}=\\
				&=\frac{\omega}{N^3}\ton*{\sum_{j,k,l}^N\xi_j^1\tilde{\xi}_j^{1}+\sum_{\mu=2}^K\sum_{j,k,l}^N\xi_i^1\xi_j^1\xi_k^1\xi_l^1\xi_i^{\mu}\tilde{\xi}_j^{\mu}\xi_k^{\mu}\xi_l^{\mu}},
			\end{align*}
			therefore 
			\[
				(b)\sim\omega\ton*{\frac{1}{\sqrt{N}}+\sqrt{\frac{K}{N^3}}}.
			\]
			Combining the two terms we get 
			\[
				R^{(1)}=(a)+3(b)\sim\omega\ton*{1+\frac{1}{\sqrt{N}}+\sqrt{\frac{K}{N^3}}}.
			\]
			We can now turn to the second order of pattern noise, proceeding as before it is easy to show that
			\[
				R^{(2)}=\frac{\omega^2}{N^3}\sum_{\mu=1}^K\sum_{j,k,l}^N\xi_i^1\xi_j^1\xi_k^1\xi_l^1\ton*{3\underbrace{\tilde{\xi}_i^{\mu}\tilde{\xi}_j^{\mu}\xi_k^{\mu}\xi_l^{\mu}}_{(a)}+3\underbrace{\xi_i^{\mu}\tilde{\xi}_j^{\mu}\tilde{\xi}_k^{\mu}\xi_l^{\mu}}_{(b)}}.
			\]
			The first term is 
			\begin{align*}
				(a)=&\frac{\omega^2}{N^3}\sum_{j,k,l}^N\xi_i^1\xi_j^1\xi_k^1\xi_l^1\tilde{\xi}_i^{1}\tilde{\xi}_j^{1}\xi_k^{1}\xi_l^{1}+\\
				&+\frac{\omega^2}{N^3}\sum_{\mu=2}^K\sum_{j,k,l}^N\xi_i^1\xi_j^1\xi_k^1\xi_l^1\tilde{\xi}_i^{\mu}\tilde{\xi}_j^{\mu}\xi_k^{\mu}\xi_l^{\mu}=\\
				&=\frac{\omega^2}{N^3}\ton*{\sum_{j,k,l}^N\xi_i^1\tilde{\xi}_i^{1}\xi_j^1\tilde{\xi}_j^{1}+\sum_{\mu=2}^K\sum_{j,k,l}^N\xi_i^1\xi_j^1\xi_k^1\xi_l^1\tilde{\xi}_i^{\mu}\tilde{\xi}_j^{\mu}\xi_k^{\mu}\xi_l^{\mu}}.
			\end{align*}
			Consequently
			\[
				(a)\sim\omega^2\ton*{\frac{1}{\sqrt{N}}+\sqrt{\frac{K}{N^3}}}.
			\]
			Analogously 
			\begin{align*}
				(b)=&\frac{\omega^2}{N^3}\sum_{j,k,l}^N\xi_i^1\xi_j^1\xi_k^1\xi_l^1\xi_i^{1}\tilde{\xi}_j^{1}\tilde{\xi}_k^{1}\xi_l^{1}+\\
				&+\frac{\omega^2}{N^3}\sum_{\mu=2}^K\sum_{j,k,l}^N\xi_i^1\xi_j^1\xi_k^1\xi_l^1\xi_i^{\mu}\tilde{\xi}_j^{\mu}\tilde{\xi}_k^{\mu}\xi_l^{\mu}=\\
				&=\frac{\omega^2}{N^3}\ton*{\sum_{j,k,l}^N\xi_k^1\tilde{\xi}_k^{1}\xi_j^1\tilde{\xi}_j^{1}+\sum_{\mu=2}^K\sum_{j,k,l}^N\xi_i^1\xi_j^1\xi_k^1\xi_l^1\xi_i^{\mu}\tilde{\xi}_j^{\mu}\tilde{\xi}_k^{\mu}\xi_l^{\mu}}.
			\end{align*}
			That is
			\[
				(b)\sim\omega^2\ton*{\frac{1}{N}+\sqrt{\frac{K}{N^3}}}.
			\]
			We then obtain
			\[
				R^{(2)}=3(a)+3(b)\sim\omega^2\ton*{\frac{1}{\sqrt{N}}+\frac{1}{N}+\sqrt{\frac{K}{N^3}}}.
			\]
			The third order of noise is of the form
			\[
				R^{(3)}=\frac{\omega^3}{N^3}\sum_{\mu=1}^K\sum_{j,k,l}^N\xi_i^1\xi_j^1\xi_k^1\xi_l^1\ton*{3\underbrace{\tilde{\xi}_i^{\mu}\tilde{\xi}_j^{\mu}\tilde{\xi}_k^{\mu}\xi_l^{\mu}}_{(a)}+\underbrace{\xi_i^{\mu}\tilde{\xi}_j^{\mu}\tilde{\xi}_k^{\mu}\tilde{\xi}_l^{\mu}}_{(b)}},
			\]
			where the two terms scale as 
			\[
				(a)\sim\omega^3\ton*{\frac{1}{N}+\sqrt{\frac{K}{N^3}}},
			\]
			\[
				(b)\sim\omega^3\ton*{\sqrt{\frac{1}{N^3}}+\sqrt{\frac{K}{N^3}}}\sim\sqrt{\frac{K}{N^3}}.
			\]
			Therefore 
			\[
				R^{(3)}=3(a)+(b)\sim\omega^3\ton*{\frac{1}{N}+\sqrt{\frac{K}{N^3}}}.
			\]
			Finally the fourth order is 
			\[
				R^{(4)}=\frac{\omega^4}{N^3}\sum_{\mu=1}^K\sum_{j,k,l}^N\xi_i^1\xi_j^1\xi_k^1\xi_l^1\tilde{\xi}_i^{\mu}\tilde{\xi}_j^{\mu}\tilde{\xi}_k^{\mu}\tilde{\xi}_l^{\mu},
			\]
			whose scaling is simply
			\[
				R^{(4)}\sim\omega^4\sqrt{\frac{K}{N^3}}.
			\]
			Combining the four contribution we obtain the following scaling for the noise
			\begin{align*}
				R=\sum_{i=0}^4R^{(i)}\sim&\omega\ton*{1+\frac{1}{\sqrt{N}}}+\omega^2\ton*{\frac{1}{\sqrt{N}}+\frac{1}{N}}+\\
				&+\omega^3\frac{1}{N}+\sqrt{\frac{K}{N^3}}\ton*{1+\omega+\omega^2+\omega^3+\omega^4}.
			\end{align*}
			Recalling that $S\sim1$ we deduce that the network can tolerate, at most, $\omega\sim1$. In other words the tolerance versus pattern noise satisfies 
			\[
				\beta(a)\sim1\ \text{for}\ a\leq3.
			\]
		\subsection{Noisy learning}
			At second level we can consider the following form of synaptic noise
			\begin{equation}
				\eta_{ij}^{\mu}=\xi_i^{\mu}\xi_j^{\mu}+\omega\tilde{\xi}_{ij}^{\mu}.
				\label{eq:noise_alemanno}
			\end{equation}
			The local field is defined as
			\[
				h_i=\frac{1}{N^3}\sum_{\mu=1}^K\sum_{j,k,l}^N\eta_{ij}^{\mu}\eta_{kl}^{\mu}\sigma_j\sigma_k\sigma_l,
			\]
			where, even if not specified, the sum does not contain self-interaction among spins. In these terms the Hamiltonian is
			\[
				H=-\sum_i^Nh_i\sigma_i.
			\]
			We want to study the stability of pattern $\xi_i^1$. Recalling that $\eta_{ij}^{\mu}=\xi_i^{\mu}\xi_j^{\mu}+\omega\tilde{\xi}_{ij}^{\mu}$ we get
			\begin{align*}
				h_i\xi_i^1&=\frac{1}{N^3}\sum_{\mu=1}^K\sum_{j,k,l}^N\eta_{ij}^{\mu}\eta_{kl}^{\mu}\xi_i^1\xi_j^1\xi_k^1\xi_l^1\\
				&=\frac{1}{N^3}\sum_{\mu=1}^K\sum_{j,k,l}^N\ton*{\xi_i^{\mu}\xi_j^{\mu}+\omega\tilde{\xi}_{ij}^{\mu}}\ton*{\xi_k^{\mu}\xi_l^{\mu}+\omega\tilde{\xi}_{kl}^{\mu}}\xi_i^1\xi_j^1\xi_k^1\xi_l^1,
			\end{align*}
			that is
			\begin{align*}
				h_i\xi_i^1=\frac{1}{N^3}\sum_{\mu=1}^K\sum_{j,k,l}^N\xi_i^1\xi_j^1\xi_k^1\xi_l^1&\left(\xi_i^{\mu}\xi_j^{\mu}\xi_k^{\mu}\xi_l^{\mu}+\omega\xi_i^{\mu}\xi_j^{\mu}\tilde{\xi}_{kl}^{\mu}+\right.\\
				&\left.+\omega\xi_k^{\mu}\xi_l^{\mu}\tilde{\xi}_{ij}^{\mu}+\omega^2\tilde{\xi}_{ij}^{\mu}\tilde{\xi}_{kl}^{\mu}\right).
			\end{align*}
			We can split this sum in signal $S$ and noise $R=R^{(0)}+R^{(1)}+R^{(2)}$. The signal is 
			\[
				S=\frac{1}{2N^3}\sum_{j,k,l}^N\xi_i^1\xi_j^1\xi_k^1\xi_l^1\xi_i^{1}\xi_j^{1}\xi_k^{1}\xi_l^{1}\sim1.
			\]
			The contribution to noise due to interference among patterns $R^{(0)}$ is 
			\[
				R^{(0)}=\frac{1}{N^3}\sum_{\mu=2}^k\sum_{j,k,l}^N\xi_i^1\xi_j^1\xi_k^1\xi_l^1\xi_i^{\mu}\xi_j^{\mu}\xi_k^{\mu}\xi_l^{\mu}\sim\frac{\sqrt{N^3K}}{N^3}\sim\sqrt{\frac{K}{N^3}}.
			\]
			As expected, in absence of pattern noise, the network can store up to $N^3$ vector patterns. At first order synaptic noise contributes with $R^{(1)}$, whose expression is 
			\[
				R^{(1)}=\frac{\omega}{N^3}\sum_{\mu=1}^K\sum_{j,k,l}^N\xi_i^1\xi_j^1\xi_k^1\xi_l^1\ton*{\xi_i^{\mu}\xi_j^{\mu}\tilde{\xi}_{kl}^{\mu}+\xi_k^{\mu}\xi_l^{\mu}\tilde{\xi}_{ij}^{\mu}}.
			\]
			Distinguishing between $\mu=1$ and $\mu>2$ we get 
			\begin{align*}
				R^{(1)}=&\frac{\omega}{N^3}\sum_{j,k,l}^N\xi_i^1\xi_j^1\xi_k^1\xi_l^1\ton*{\xi_i^{1}\xi_j^{1}\tilde{\xi}_{kl}^{1}+\xi_k^{1}\xi_l^{1}\tilde{\xi}_{ij}^{1}}+\\
				&+\frac{\omega}{N^3}\sum_{\mu>2}^K\sum_{j,k,l}^N\xi_i^1\xi_j^1\xi_k^1\xi_l^1\ton*{\xi_i^{\mu}\xi_j^{\mu}\tilde{\xi}_{kl}^{\mu}+\xi_k^{\mu}\xi_l^{\mu}\tilde{\xi}_{ij}^{\mu}}=\\
				=&\frac{\omega}{N^3}\sum_{j,k,l}^N\ton*{\xi_k^1\xi_l^1\tilde{\xi}_{kl}^{1}+\xi_i^{1}\xi_j^{1}\tilde{\xi}_{ij}^{1}}+\\
				&+\frac{\omega}{N^3}\sum_{\mu>2}^K\sum_{j,k,l}^N\xi_i^1\xi_j^1\xi_k^1\xi_l^1\ton*{\xi_i^{\mu}\xi_j^{\mu}\tilde{\xi}_{kl}^{\mu}+\xi_k^{\mu}\xi_l^{\mu}\tilde{\xi}_{ij}^{\mu}}.
			\end{align*}
			We then obtain
			\begin{align*}
				R^{(1)}&\sim\frac{\omega}{N^3}\ton*{N\sqrt{N^2}+N^2\sqrt{N}+\sqrt{KN^3}}\\
				&\sim\omega\ton*{\frac{1}{N}+\frac{1}{\sqrt{N}}+\sqrt{\frac{K}{N^3}}}.
			\end{align*}
			Finally the second order of the pattern noise $R^{(2)}$ is
			\[
				R^{(2)}=\frac{\omega^2}{N^3}\sum_{\mu}^K\sum_{j,k,l}^N\xi_i^1\xi_j^1\xi_k^1\xi_l^1\tilde{\xi}_{ij}^{\mu}\tilde{\xi}_{kl}^{\mu}\sim\omega^2\sqrt{\frac{K}{N^3}}.
			\]
			In conclusion the noise can be written as
			\begin{align*}
				R&\sim R^{(0)}+R^{(1)}+R^{(2)}\sim\\
				&\sim\sqrt{\frac{K}{N^3}}\ton*{1+\omega+\omega^2}+\omega\ton*{\frac{1}{N}+\frac{1}{\sqrt{N}}}.
			\end{align*}
			We set $K\sim N^a$ and $\omega\sim N^b$, in this way we obtain, at leading order 
			\[
				R\sim\sqrt{\frac{K}{N^3}}\omega^2+\frac{\omega}{\sqrt{N}}\sim N^{(a-3)/2+2b}+N^{b-1/2}.
			\]
			Recalling that retrieval is possible provided that $R\lesssim S\sim1$ we see that there are two different regimes
			\begin{itemize}
				\item $\mathbf{a\leq1}$\\
				In this case noise in dominated by the second term and the extremal condition for retrieval reads 
				\[
					N^{b-1/2}\sim1\to b=\frac{1}{2}.
				\]
				Therefore the tolerance versus pattern noise is
				\[
					\beta(a)\sim N^{1/2} \ \text{for}\ a\leq1.
				\]
				\item $\mathbf{a>1}$\\
				Increasing the load reduces the tolerance versus pattern noise, indeed we obtain 
				\[
					N^{(a-3)/2+2b}\sim1\to b=\frac{3}{4}-\frac{a}{4}.
				\]
				It then follows 
				\[
					\beta(a)\sim N^{(3-a)/4} \ \text{for}\ 1<a<3.
				\]
			\end{itemize}
		\subsection{Noisy storing}
			Finally the less challenging noise is the one applied on 4-tensors or, analogously, on the couplings. This is of the form
			\begin{equation}
				\eta_{ijkl}^{\mu}=\xi_{i}^{\mu}\xi_{j}^{\mu}\xi_{k}^{\mu}\xi_{l}^{\mu}+\omega\tilde{\xi}_{ijkl}^{\mu}.
				\label{eq:noise_third_level}
			\end{equation}
			Again we consider the product between the local field $h_i$ and $\xi_i^1$
			\[
				h_i\xi_i^1=\frac{1}{2N^3}\sum_{\mu=1}^K\sum_{j,k,l}^N\ton*{\xi_i^{\mu}\xi_j^{\mu}\xi_k^{\mu}\xi_l^{\mu}+\omega\tilde{\xi}_{ijkl}^{\mu}}\xi_i^1\xi_j^1\xi_k^1\xi_l^1.
			\]
			The signal, as already shown, scales as $S\sim1$, while the noise is composed of two contributions: zeroth and first order. We have already computed the former
			\[
				R^{(0)}\sim\sqrt{\frac{K}{N^3}}\sim\sqrt{\frac{K}{N^3}}.
			\]
			For what concerns the first order it holds 
			\[
				R^{(1)}=\frac{\omega}{2N^3}\sum_{\mu=1}^K\sum_{j,k,l}^N\tilde{\xi}_{ijkl}^{\mu}\xi_i^1\xi_j^1\xi_k^1\xi_l^1\sim\omega\sqrt{\frac{K}{N^3}}.
			\]
			Therefore 
			\[
				R=R^{(0)}+R^{(1)}\sim\sqrt{\frac{K}{N^3}}\ton*{1+\omega}\sim\omega\sqrt{\frac{K}{N^3}}.
			\]
			Setting, as before, $K\sim N^a$ and $\omega\sim N^b$ the condition for retrieval becomes
			\[
				N^{(a-3)/2+b}\sim1\to b=\frac{3-a}{2},
			\]
			which implies that the tolerance versus pattern noise is
			\[
				\beta(a)\sim N^{(3-a)/2}\ \text{for}\ a\leq3.
			\]

		\bibliographystyle{unsrt}

\end{document}